\documentstyle[psfig]{l-aa}

\begin{document}
\newcommand{\PSR}{PSR J0218$+$4232} 

\newcommand{\gtap}{\mathrel{\hbox{\rlap{\lower.55ex \hbox {$\sim$}}
                   \kern-.3em \raise.4ex \hbox{$>$}}}}
\newcommand{\ltap}{\mathrel{\hbox{\rlap{\lower.55ex \hbox {$\sim$}}
                   \kern-.3em \raise.4ex \hbox{$<$}}}}
\thesaurus{08.14.1; 08.16.7 PSR~J0218$+$4232; 13.25.5}
\title{The pulse shape and spectrum of the millisecond pulsar PSR J0218$+$4232 
in the energy band 1-10 keV observed with BeppoSAX}
\author{T.~Mineo\inst{1}
\and G.~Cusumano\inst{1}
\and L.~Kuiper\inst{2}
\and W.~Hermsen\inst{2}
\and E.~Massaro\inst{3}
\and W.~Becker\inst{4}
\and L.~Nicastro\inst{1}
\and B.~Sacco\inst{1}
\and F.~Verbunt\inst{5}
\and A.G.~Lyne\inst{6}
\and I.H.~Stairs\inst{6}
\and S.~Shibata\inst{7}
}
\institute{Istituto di Fisica Cosmica ed Applicazioni
all'Informatica CNR, Via U. La Malfa 153, I-90146, Palermo, Italy \and
SRON-Utrecht, Sorbonnelaan 2, 3584 CA Utrecht, The Netherlands \and
Istituto Astronomico, Unita' GIFCO Roma-1, Universita' "La Sapienza",
Via Lancisi 29, I-00161, Roma, Italy \and
Max-Planck-Institut f\"{u}r Extraterrestrische Physik, D-85740
Garching bei M\"{u}nchen, Germany \and
Astronomical Institute Utrecht University, P.O. Box 80000, 3508 TA Utrecht,
 The Netherlands \and
University of Manchester, Jodrell Bank, Macclesfield SK11 9DL, 
United Kingdom \and
Department of Physics, Yamagata University, Kojirakawa, 
Yamagata 990-8560, Japan }

\offprints{T. Mineo: mineo@ifcai.pa.cnr.it}
\date{Received ....; accepted ....}
\maketitle
\markboth{T. Mineo et al.: Pulse shape and spectrum of PSR J0218+4232
observed with BeppoSAX}
{T. Mineo et al.: Pulse shape and spectrum of PSR J0218+4232 observed 
with BeppoSAX}

\begin{abstract}

We present the results of a BeppoSAX observation of \PSR\ which, for 
the first time, provides detailed information on the pulsar's temporal 
and spectral emission properties in the broad band   1--10 keV.    
We detected X-ray pulses with a pulsed fraction
of 73$\pm$12 \%. The pulse profile is characterized by two peaks phase separated 
by $\Delta\phi=0.47\pm0.05$.
The pulsed spectrum is best described by a power--law of photon index 
0.61$\pm$0.32 with an
unabsorbed (2--10 keV) X-ray flux of 4.1$\times$10$^{-13}$ 
erg cm$^{-2}$ s$^{-1}$
 implying a luminosity
of $L_x = 1.3\times 10^{32}  \,\Theta \; (d/5.7\;{\rm kpc})^2 $ erg s$^{-1}$ 
and
an X-ray efficiency of $L_x/\dot{E}=4.8\times 10^{-4} 
 \, \Theta \; (d/5.7\;{\rm kpc})^2 $  
where $\Theta$ is the solid angle spanned by the emission beam.
\end{abstract}

\section{Introduction}

PSR J0218$+$4232 is a  2.3 ms pulsar in a two day 
orbit around a $\sim$ 0.2 M$_\odot$ white dwarf companion  (Navarro 
et al.~1995) with a period derivative of $\dot{P}=8\times 10^{-20}$ 
s s$^{-1}$.
The pulsar has a spin-down energy of $\dot{E}=2.5\times 10^{35}$ erg s$^{-1}$,
a dipolar magnetic field component at the star surface of 
$B_\perp=4.3\times 10^8$ G and a 
spin-down age of $\le 4.6\times 10^8$ years. The pulsar distance inferred
from its dispersion measure and from the electron 
density model of Taylor \& Cordes (1993) is $\ge 5.7$ kpc.  
A detection of the companion star at optical wavelength was reported 
recently by Van Kerkwijk (1996). 

Soft X-ray emission from the pulsar was first detected by Verbunt
et al.~(1996) using the ROSAT HRI. A follow-up observation confirmed 
the detection and discovered  X-ray pulsation at  a significance of 
about 5 $\sigma$ using 200 HRI counts (Kuiper et al.~1998). 
The X-ray pulse profile deduced from those data is characterized by a 
sharp main pulse with an indication of a second peak at a 
phase separation of $\Delta\phi \sim 0.47$. The pulsed fraction 
inferred from
the ROSAT HRI data is 37$\pm$13 \%. 
Furthermore, Kuiper et al.~(1998) show that the measured large DC component is
consistent with a $\sim$  14\arcsec \, diameter  compact nebula surrounding the pulsar,
 but 
confirmation is required. However, it is interesting to note that also 
in the 
radio domain the source exhibits an unusually high DC component of 
$\sim$ 50 \% (Navarro et al.~1995).
The HRI provides no spectral information and the number of counts
recorded in a serendipitous off--axis PSPC observation does not allow spectral
modeling.
Also ASCA detected this source, 
however, the observation was too short: no pulsation could be seen, and a 
spectral fit  with a power--law 
photon index of 1.6$\pm$0.6   could 
only be made to the weak total excess (Kawai \& Saito 1999). 
Therefore, no detailed spectral information on the pulsed X-ray emission was 
available prior to the BeppoSAX observation reported in this paper
(see Becker \& Tr\"umper 1999 for a review of the X-ray properties of millisecond
pulsars).

Noting the spatial coincidence of PSR J0218+4232 with the EGRET source
2EG J0220§+§4228, Verbunt et al. (1996) tentatively identified the pulsar with
the high-energy $\gamma$-ray source. Using some additional EGRET observations, and
applying a combination of spatial and timing analysis, Kuiper et al. (1999)
conclude that 2EG J0220$+$4228 is probably a multiple source: between 0.1 and 1 GeV
PSR J0218$+$4232 is the most likely counterpart, and above 1 GeV the bright BL Lac 3C 66A is 
the best candidate counterpart. The third EGRET catalog (Hartman 
et al. 1999), which is based on more viewing periods than the 2EG catalog, 
identifies 3EG J0222$+$4253 (2EG J0220$+$4228) with 3C 66A, rather than 
with the ms-pulsar.
However, in a note on this source, they indicate that the identification with
3C 66A stems from  the catalog position based on the $>$ 1 GeV map. Furthermore, 
they confirm that for lower energies (100-300 MeV) the EGRET map is
consistent with all the source flux coming from the pulsar, 3C 66A being 
statistically excluded.

In this paper we present the results of a  BeppoSAX observation 
of \PSR\
which, for the first time, provides detailed information on the pulsar's
temporal and spectral emission properties across a wide band from 1 to 10 keV.

\section{Observation}

\PSR\ was observed on January 1999 14--16 by the Narrow Field
Instruments (NFIs) aboard BeppoSAX satellite (Boella et al. 1997a).
We report in this paper 
only results from data collected with the MECS instrument 
(Boella et al. 1997b), sensitive in
the energy range 1-10 keV; the net exposure time was  82\,795 s.
The LECS instrument (0.1--10 keV; Parmar et al. 1997)
observed the source for a much shorter time due to the  constraint
on operating  only  during spacecraft night.
The HPGSPC (4--60 keV; Manzo et al. 1997) was not working during the observation. 
A signal detected with the PDS (13--200 keV; Frontera et al. 1997) will
  be shortly discussed in Section 5.
No pulsed signals have been detected in the LECS and PDS data.

During the observation only two of the three MECS detectors were operating
(MECS2 and MECS3).
MECS has a field of view of $56'$ and an angular resolution of about 
1\farcm2 at 6 keV.
During the observation of \PSR\ the instrument operated in direct
mode, transmitting to ground information on each individual photon.
Standard procedures and selection criteria were applied on the observation
data to avoid the South Atlantic Anomaly, solar, bright Earth and
particle contaminations\footnote{see http://www.sdc.asi.it/software/cookbook 
as a reference about the data analysis software and reduction procedures.}.
Event reduction has been performed using the SAXDAS v.2.0.0 package.

\section{Spatial analysis}

A 100 ks ROSAT HRI (0.1--2.4 keV) observation of \PSR\ (see Kuiper et
al. 1998) revealed 7  X-ray sources within 
a radius of $\sim5^{\prime}$ around our target (see Fig. 1).
Given the extended tails of the MECS Point-Spread Function (PSF),
accurate spatial analysis is required to separate the signal of  
\PSR\ from that of any neighbouring sources. We used a Maximum Likelihood 
approach in searching for individual sources 
on top of a background model, as well as for analyzing simultaneously several 
sources and a background model.
\\
The search is applied on various equidistant trial positions within a 
selected part of the instrument field of view, where
it checks for the presence of a source taking into account the 
Poissonian nature of the data. 
Maximum Likelihood Ratio (MLR) tests are performed describing in the zero 
{${\cal{H}}_0$} hypothesis the 2-d
event distribution in terms of a flat background model only, while in the 
alternative
{${\cal{H}}_1$} hypothesis the data is described in terms of a point source 
at the trial position and a flat background model (see e.g. Kuiper 
et al. 1998 for a more detailed description).

The improvement in the likelihood {$\cal{L}$} between 
{${\cal{H}}_0$} and {${\cal{H}}_1$} computed from the quantity 
$Q=-2\ln({\cal{L}}^{H_0}/{\cal{L}}^{H_1})$ 
yields simultaneously the detection significance and strength of the source. 
The probability distribution of $Q$ is that of a $\chi^2$ for ${n_1}-{n_0}$ 
degree of freedom (d.o.f.) where ${n_0}$ and ${n_1}$ are the d.o.f. 
for the {${\cal{H}}_0$} and  {${\cal{H}}_1$} hypotheses, respectively.
In searching for sources  ${n_1}-{n_0}$ = 3, while in a detection test of 
a known source ${n_1}-{n_0}$ = 1. The distribution of Q as a function of 
trial position is the MLR--map.

\begin{figure*}
\centerline{
\hbox{
\psfig{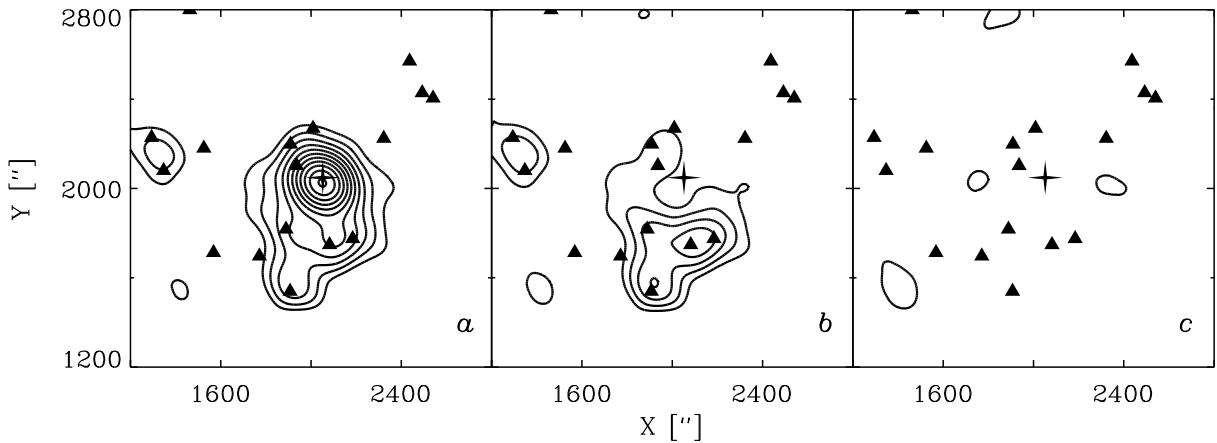}
}}
\caption{
Maximum Likelihood Ratio maps in instrumental coordinates ($X[^{\prime\prime}], 
Y[^{\prime\prime}]$) using all MECS data 
($\sim$ 1--10 keV)
for the sky region around PSR J0218+4232 (star). The contours start at a value 
of 3 $\sigma$ source detection significance with a
stepsize of 2 $\sigma$ (for 1 d.o.f.) or equivalently at $Q$ values (see text) of
$9,25,49,\dots$. The positions of all sources detected by the ROSAT HRI 
observation below 2.4 keV are indicated by filled triangles. $a$) Evidence for 
sources above a flat background showing PSR J0218+4232 ($\sim 21 \, \sigma$) on top
of an extended excess. $b$) Like $a$), but PSR J0218+4232 has now been
``subtracted''. $c$) No significant excesses are left when 9 ROSAT sources are
added to the background model
}
\end{figure*}

We applied the above approach to the MECS data without energy selection
allowing for maximum source statistics. 
We used an energy averaged PSF assuming  a power--law spectral 
shape of index $1.5$, given by the following expression:

$$PSF(x,y) = $$
$$\frac{\int_{0}^{\infty} E^{-1.5} \cdot (S_{2}(E)+S_{3}(E)) \cdot 
PSF(x,y\ \vert\ E)  \cdot  dE \ } {\int_{0}^{\infty} E^{-1.5} \cdot  (S_{2}(E) + 
S_{3}(E))  
\cdot  dE}, \eqno(1)$$
where $S_{2}(E)$ and $S_{3}(E)$ are  the sensitive
areas of the two operating MECS units.
The $PSF(x,y \ \vert \ E)$ is modelled using the parametric form given 
in  Boella
et al. (1997b). Furthermore, we investigated the impact of the 
assumed power--law index on the MLR results and found that this is 
negligible, as expected by the moderate 
dependence
on energy of the PSF.

The resulting MLR map/image is shown in Fig. 1a. The maximum value in this 
figure ($ > 21 \,\sigma$) is reached at a position  consistent with 
that determined by
the ROSAT HRI for  \PSR\
(indicated by a star symbol). 
 It is, however, evident that several of
the sources detected by ROSAT below 2.4 keV (indicated by filled triangles)
contribute significantly to the extended excess.
We can show this better, by repeating the search for sources on top of a flat
background model {\em and} a point source at the pulsar position 
(effectively ``subtracting'' the counts from \PSR\ from Fig. 1a).
The remaining extended feature (Fig. 1b) clearly follows the distribution of 
the nearby sources, reaching  detection significances up to about 10~$\sigma$. 
Figure 1c shows the MLR map resulting from a point
source search on top of a flat background and 9  sources fixed at the 
ROSAT HRI source positions: no significant residual emission remains.
The total number of counts assigned by this analysis to \PSR\ is 403$\pm$29.

The position of \PSR\ determined from the MLR map shown in Fig. 1a 
is shifted by  $\sim$ 23\arcsec \,with respect to the radio position, 
well within the systematic uncertainty of the accuracy of the BeppoSAX 
pointing 
reconstruction
\footnote{see http://www.sdc.asi.it/software/cookbook/attitude.html}.
\\
The spatial analysis allows also the evaluation of an optimum event
 extraction radius 
by computing the Signal-to-Noise ratio as a function of radial distance 
$r$ from the source position. Taking as source radial profile the PSF,
and as noise the {\em measured} radial count distribution, 
we obtained the value $r = 100^{\prime\prime}$. Note that none of the nearby sources 
fall within this radius. 
The selected region contains 62 \% (250 counts) of the MECS source signal;
the number of background
events in the same area is 92 with a  contribution
from the neighbouring sources of 12 counts.

\section{Timing analysis}

The arrival times of all selected events were converted to the Solar System 
Barycentric Frame using BARYCONV\footnote{see 
http://www.sdc.asi.it/software/saxdas/baryconv.html},  then
folded according to the radio ephemeris 
(see Table 1),  correcting for the pulsar binary motion. 
The pulse phase distribution deviates from a statistically flat distribution
at a 6.8 $\sigma$ level applying a $Z^{2}$ test (Buccheri et al. 1983)
using the first two harmonics.

The MECS timing resolution is dominated by the precision of the 
OBT~(On Board Time)-UTC 
conversion: an inspection of the residuals of the OBT-UTC linear fit 
reveals a systematic scatter with an rms  of the order of $\sim$ 0.2 ms 
that cannot be reduced
by fitting higher order polynomials. This effect 
corresponds to an uncertainty in the photon phases of 0.087.

\begin{table}
\caption{Ephemeris of PSR J0218$+$4232}
\begin{flushleft}
\begin{tabular}{ll}
\hline\noalign{\smallskip}
\noalign{\smallskip}
Parameter &  Value  \\
\hline
Right Ascension (J2000) &  02$^{\rm h}$ 18$^{\rm m}$ 6\fs350    \\
Declination (J2000) &  42$^\circ$ $32'$ 17\farcs44     \\
Epoch validity start/end (MJD) &  49092 -- 50900\\
Frequency &  430.4610674213 Hz\\
Frequency derivative &  $-1.4342\times 10^{-14}$ Hz s$^{-1}$ \\
Epoch of the period (MJD) &  49996.000000023 \\
Orbital period &  175292.3020 s    \\
a$_p$sin i  &  1.98444 (lt-s) \\
Eccentricity &  0  \\
Longitude of periastron &  0 \\
Time of ascending node (MJD) &  49996.637640  \\
\noalign{\smallskip}
\hline
\end{tabular}
\end{flushleft}
\end{table}

\begin{figure}
\centerline{
\vbox{
\psfig{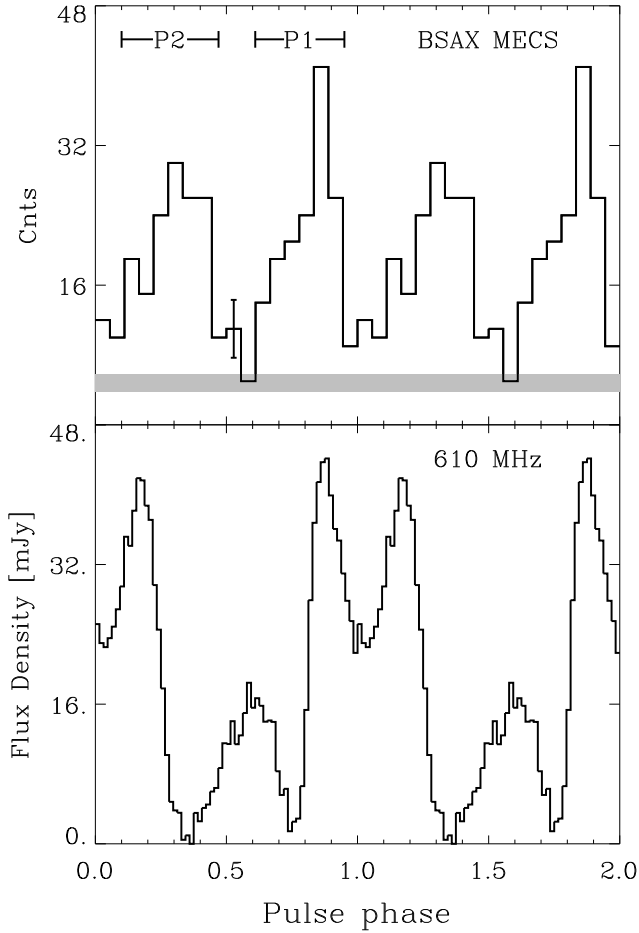}
}} 
\caption{PSR J0218$+$4232 light curve in the 1--10 keV range as observed
by the MECS (top panel) and at 610 MHz (bottom panel, Stairs et al. 1999). 
In the top panel  a typical statistical 1 $\sigma$ error bar
is shown and the MECS background level as determined in the spatial analysis is 
indicated as a grey band, representing the $\pm$1 $\sigma$  widths. The two
horizontal bars above the MECS profile indicate the definitions of the phase
intervals for the two peaks, Peak 1 (P1) and Peak 2 (P2), as explained in the 
text.
The alignment between the two curves is arbitrary  
}
\end{figure}

\noindent

The light curve resulting from  folding  all MECS events (1--10 keV) 
with a phase resolution of 18 bins ($\sim$ 0.13 ms)
is shown in Fig. 2 (top panel): two cycles are plotted for clarity. 
The MECS background level, indicated with a grey band ($\pm$1 $\sigma$ errors),
is determined in the spatial analysis and therefore includes
 the contributions from the neighbouring sources.
 In the same figure (bottom panel) the 610 MHz profile as derived by Stairs et al.
(1999) is also presented. The phase relation between the two profiles 
is unknown for the time being: they have been arbitrarily aligned.
Notice that, like in the radio light curve, the MECS profile is not simple:
it has a prominent double peak structure with a relative phase separation (centre 
to centre) of 0.47$\pm$0.05, confirming the value derived by Kuiper et al. (1998)
from ROSAT data below 2.4 keV.
\\
The phase histograms for 1.6--4 keV and 4--10 keV are shown respectively in 
the middle and bottom panels of Fig. 3. The ROSAT (0.1--2.4 keV) profile
(Kuiper et al. 1998),
shifted in phase to obtain the highest peak coincident with the most 
significant one in the MECS softer light curve (middle panel), 
is also shown for
comparison in the top panel of Fig. 3. These profiles clearly show  a change of the
relative peak intensities. 
The peak at phase 0.8 is stronger in the low energy histograms
(top and middle panels), 
while that at phase 0.3 is more prominent at high energies (bottom panel). 

The structure measured in the MECS light curves of two peaks separated 
by  narrow valleys and the uncertainty due to the OBT--UTC conversion 
residuals make it difficult to establish an unpulsed interval:
slightly different values could affect the unpulsed level. 
Notice, moreover, that while in the ROSAT profile a DC component is  
apparent above the background level, the same does not hold for the
4--10 keV profile, where  the 
background level is consistent with those measured in the valleys of the 
light curve.
In the intermediate energy profile, at face value, there might be
some evidence of a DC component but not as high as in the ROSAT profile.
Such effect could be due to the presence of a quite soft unpulsed emission.
In an attempt to 
quantify the evidence for a DC component, we applied the bootstrap method 
proposed by Swanepoel et al. (1996), which allows to estimate the DC level, 
and therefore also the pulsed fraction, working directly on the phases of
individual photons. However, in the
available form, this method is able to find only one unpulsed interval and cannot 
account for systematic errors. 
One should therefore realize that the quoted errors on the calculated 
parameters are  only statistical. 
Using this algorithm we found, for the whole
MECS energy band, an unpulsed interval of $0.47$--$0.61$ and a corresponding
pulsed fraction of 0.73$\pm$0.12 (1~$\sigma$ error).
Notice that the average count level in the interval $-0.08$--$0.15$ is
statistically
not different from that measured in the former interval. 
The resulting DC--fractions for the three bands of Fig. 3 are
 shown in Fig. 4.
 The decrease with energy is clearly visible, but the error bars are
large and the trend deviates only at the 2.4 $\sigma$  level from a flat 
distribution. Given the large error bars and the systematic uncertainties,
we did not derive a detailed spectrum of the DC component.

\begin{figure}
\centerline{
\hbox{
\psfig{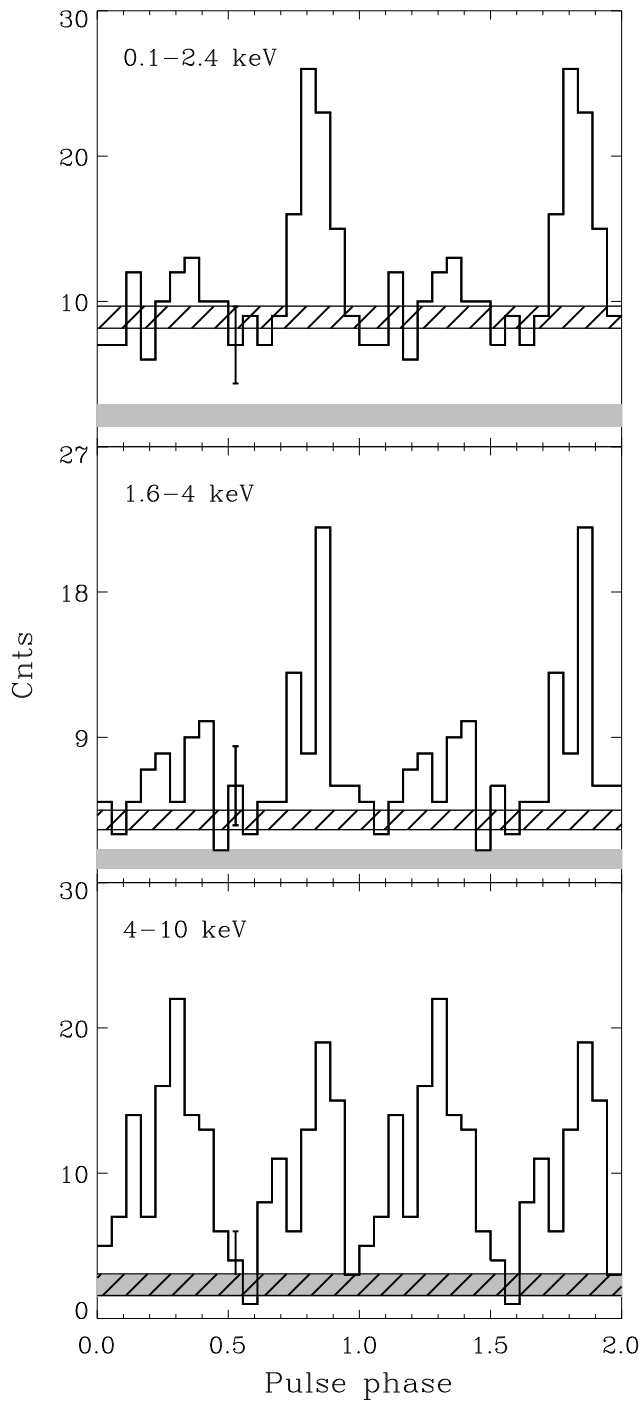}
}
}
\caption{
PSR J0218+4232 phase histograms: energy range 0.1-2.4 keV from the 
ROSAT HRI (Kuiper et al. 1998); energy ranges 1.6-4 keV and 4-10 keV from the
BeppoSAX MECS (this work). Note the increase of the P2 peak strength with
respect to P1 with  energy. 
The hatched  area represents the unpulsed  level ($\pm$1 $\sigma$)
determined in a bootstraping analysis 
of the phase histogram, while the 
grey area represents the  background level with 1 $\sigma$ errors 
determined in the spatial analysis}
\end{figure}

The relatively high scatter in the residuals of our timing solution
(up to $\sim$0.2 ms) does not allow to determine accurately the intrinsic widths 
of the two pulses. The total scatter in the event arrival times at the SSB might even be 
consistent with the measured width of P1 in Figure 2.

We defined the phase intervals containing the two pulses as follows:
Peak 1 (P1) phase interval 0.61-0.95 and Peak 2 (P2) 0.10-0.47, where P1
is identified with the obvious ROSAT peak and appears to have a softer 
spectrum than P2. The statistical significance of the suggested change in 
morphology of the phase histogram with  energy can be evaluated
by the P2/P1 ratio.
This ratio, computed in the energy bands 
1.6--4 keV and 4--10 keV  after background subtraction, is shown in 
Fig. 5. The ROSAT point derived from the data of Kuiper et al.
(1998), indicated by a filled circle, is evaluated using the same phase
definitions.
Including the ROSAT point, a correlation of P2/P1 with energy
is evident, but the statistical significance is only 2 $\sigma$,
and  confirmation is needed.

\begin{figure}
\centerline{
\vbox{
\psfig{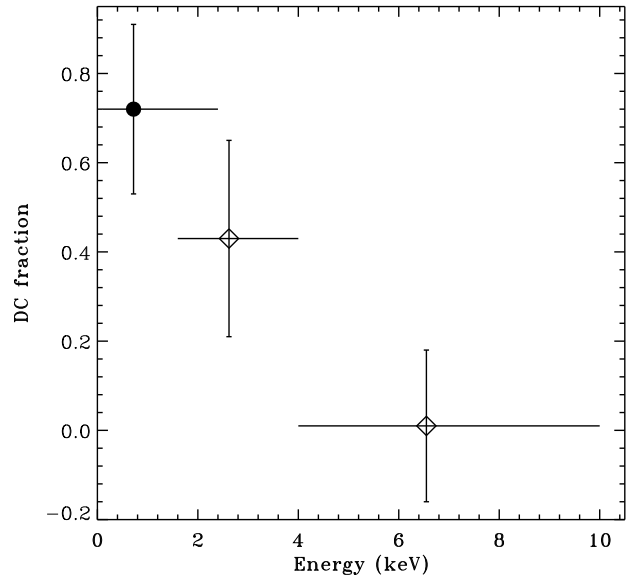}
}} 
\caption{ The DC--fraction as a function of energy.
The ROSAT point (filled circle) was derived using the same width 
of the unpulsed
interval as obtained for the whole MECS range
}
\end{figure}

\section{Spectral analysis}

Spectral fits were performed on the background subtracted MECS data binned
logarithmically after checking that  
each single energy channel contains at least 20 photons.  
The N$_H$ was fixed to the value of 5$\times$10$^{20}$ cm$^{-2}$ 
(see Verbunt et al. 1996). 
We fitted the total pulsed spectrum, i.e. the excess counts above the unpulsed 
level in the phase histogram, with a power--law model, taking into account
the energy dependent interstellar absorption.
The resulting photon 
index is  $\alpha$ = 0.61$\pm$0.32,
with a 2--10 keV flux $F=4.1\times 10^{-13}$ erg cm$^{-2}$ s$^{-1}$, 
and a reduced $\chi^2_r$ of 0.7 (3 dof).
 A fit with a black body spectral 
distribution gives $kT=2.3 \pm 0.5$ keV, with a $\chi^2_r=1.4$ (3 dof).
Both the $\chi^2$ values are acceptable, the black body temperature 
$T=2.6\times 10^7$ K, however, is quite higher than the values  measured for the 
thermal components of other pulsars that typically does not exceed $\sim 5
\times 10^{6}$ K (Greiveldinger et al. 1996,  Becker \& Tr\"umper 
 1997, 1999). 
Fig. 6 also shows the spectra for the pulses P1 and P2 separately,
using the same energy bins as for the total pulsed spectrum. We fitted again 
power-law models for each of the spectra, but the low statistics result in 
large errors in the estimated parameters. 
The measured spectral indices of P1 and P2 are $0.84\pm 0.35$
and $0.42\pm0.36$ respectively, in agreement with the trend seen in
the P2/P1 ratio.

Finally, we  determined the total spectrum (pulsed plus DC) applying 
the Maximum
Likelihood approach to the spatial distributions in smaller energy intervals
(see also Fig. 6). 
The fit with a power--law  gives a photon 
spectral index $\alpha=0.94 \pm 0.22$ with a 2--10 keV 
flux $F=4.3\times 10^{-13}$ erg cm$^{-2}$ s$^{-1}$, about 10 \%
higher than the pulsed value given above,
and a 
reduced $\chi^2_r$ of 0.56 (8 dof).
\\ 
Given the hard spectra found above,
we also analysed the PDS data to investigate whether there was any signal
from \PSR\ at higher energies, even though an extrapolation 
of the total spectrum measured in the MECS to the PDS range  predicts
a flux well below the PDS sensitivity threshold. Indeed, no pulsed signal
 was found, but in the 17-25 keV band a 
DC signal was detected at the $\sim$4 $\sigma$ level. 
Moreover the fit of  MECS+PDS spectra with a power--law plus a 
constant factor to take into account the intercalibration uncertainties
between the two instruments  leads to a spectral index of
1.4 and to a value for the constant $\sim$8, well above the expected range
of variability (0.7-1, Cusumano et al. 2000).
 Furthermore, the PDS light curve shows that most 
of the source counts are concentrated in the first half of the observation.
Therefore, it is likely that we observed a variable source in the large
field of view of the PDS collimators ($1\fdg3\times1\fdg3$ FWHM). 
Note that there are
 several sources  within a $3^\circ$ region in the soft  X-ray
catalogs, but none in the hard X-ray  ones.

\section{Discussion}

The BeppoSAX observation of \PSR\  provided the first detection of
pulsed emission from this millisecond pulsar for energies up to 
 10 keV. The source shows  a
double peaked pulse shape with a remarkably flat (photon index $\sim$0.6) spectral
distribution and indications that the peak intensity ratio depends on photon
energy.
In particular, the P2 peak, which is not prominent in the ROSAT low energy
phase histogram, becomes the dominant feature above 5 keV. 

Our fit for a black body spectrum gives the quite high value of 2.6 $\times$
10$^7$ K for the temperature of the emitting region. If we assume that
the radiation comes from the pulsar polar cap, heated, for instance, 
by the interactions with starward moving high energy particles, we can 
compute the expected luminosity. The polar cap area of \PSR, 
defined as usual by the open field lines, is of the order of 
$ A \simeq 2 \pi \, R^2 [1 - \sqrt{(1-R \,\Omega/c)}] $ = 2.9 $\times$ 10$^{11}$ cm$^2$, 
where $R$ is the stellar radius;  from the Stefan's law we obtain a bolometric luminosity 
of 5 $\times$ 10$^{36}$ erg s$^{-1}$,
a value that must be doubled if the other cap is taken into account.
Such luminosity is several orders of magnitude higher than that derived 
from the phase averaged flux 
$L_x = 1.3\times 10^{32}  \,\Theta \; (d/5.7\;{\rm kpc})^2 $ erg s$^{-1}$
(corresponding to
an X-ray efficiency of $L_x/\dot{E}=4.8\times 10^{-4} 
 \, \Theta \; (d/5.7\;{\rm kpc})^2 $)  
where $\Theta$ is the solid angle spanned by the emission beam.
It is moreover significantly greater than the spin-down energy 
loss rate. This inconsistency could be solved with the assumption that
the emission spot covers a very small fraction, of the order of about 
10$^{-4}$ or even less, of the whole cap. The flux of heating particles
should be then collimated within a very narrow angle of $\sim$ 0$^0$.2
or smaller. Furthermore, the heating flux into the two polar caps must be
different to explain the spectral difference of the two peaks.
A non--thermal origin of the X-ray emission seems therefore more convincing.

The variation of the peak ratio with energy recalls a similar effect 
observed for the Crab pulsar (see, for instance, the recent BeppoSAX data
described by Mineo et al. 1997). The possibility of detecting such
effect could be due to a particular orientation among the rotation and
magnetic axes and the line of sight. The Crab has an orthogonal alignment,
however, Navarro et al. (1995), on the basis
of the radio pulse shape, inferred for PSR J0218$+$4232 a nearly aligned rotator. 
The same
conclusion is reached by Stairs et al. (1999) who analyzed high-precision,
coherently-dispersed polarization profiles at the two frequencies of 410
and 610 MHz. Their Rotating Vector Model (RVM) fits support the 
classification of PSR J0218$+$4232 as a nearly aligned rotator with magnetic
inclination consistent with 0$^0$, namely (8$\pm$11) deg at 410 MHz and 
(8$\pm$15) at 610 MHz. Unfortunately, the impact angle parameters for 
their RVM fits have large uncertainties and therefore the line-of-sight
inclination is unconstrained.

\begin{figure}
\centerline{
\hbox{
\psfig{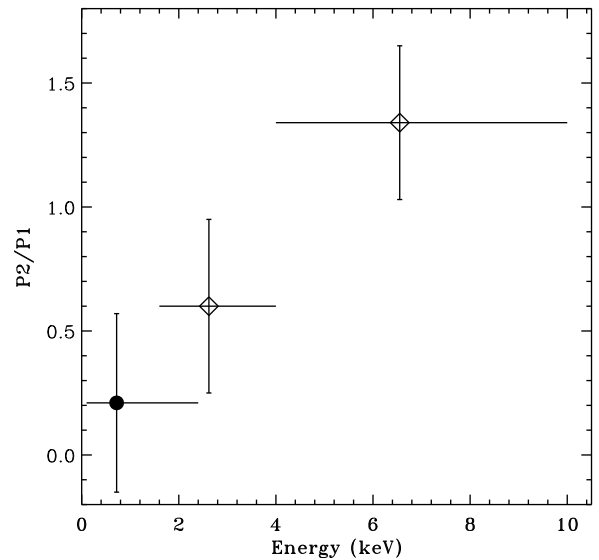}
}}
\caption{The energy dependence of P2/P1, according to the phase interval 
defined in the text. The ROSAT point is indicated by a filled circle
}
\end{figure}

The Crab-like double pulse X-ray profile of PSR J0218+4232 and its hard 
non-thermal spectrum suggests a common magnetospheric origin of the high-energy
emission.
The comparison with Crab is additionally enforced by the similar values of the
magnetic field at the light cylinder, as already pointed out by Kuiper et 
al. (1998) and Kawai \& Saito (1999). This follows directly from the classical 
assumption that the pulsar braking is entirely due to the Poynting flux at the 
light cylinder, independently of the actual structure of the magnetic field in the 
magnetosphere:

$$B_L \simeq 4 \pi^2 (\frac{I}{c^3})^{\frac{1}{2}} P^{-\frac{5}{2}} 
\dot{P}^{\frac{1}{2}}, \eqno(2)  $$

\noindent
where $I$ is the moment of inertia. For the two pulsars we obtain then:

$$B_L = B_{L,Cr} (\frac{I}{I_{Cr}})^{\frac{1}{2}} (\frac{P}{P_{Cr}})^{-\frac{5}{2}}
 (\frac{\dot{P}}{\dot{P}_{Cr}})^{\frac{1}{2}}, \eqno(3)  $$
 
\noindent
where the suffix $Cr$ is referred to Crab. Assuming that the moments of inertia 
of the two pulsars are equal, we found $B_L \simeq \frac{1}{3}B_{L,Cr} 
\simeq 3 \times
10^5$ Gauss.

The nature of the non-thermal mechanism responsible for the X-ray emission
is still unclear. A possibility is that X-ray photons are produced by
curvature radiation from relativistic electrons moving outwards along the
field lines in proximity of the light cylinder. The Lorentz factor of these
particles to radiate 
photons of frequency $\nu$ is given by 

$$\gamma = (\frac{4\pi}{3} \frac{\nu \rho}{c})^{\frac{1}{3}}, \eqno(4) $$

\noindent
where $\rho \simeq c/(3 \Omega)$ is the curvature radius of the last closed line
near  the light cylinder in a dipole field. For $\nu = 10^{18}$ Hz,
corresponding to $\sim$ 4 keV, we have:

$$\gamma \simeq (\frac{2}{9} \nu P)^{\frac{1}{3}} = 8 \times 10^4. \eqno(5) $$

\noindent
The radiative life time of the electrons is long enough to allow them to
reach the light cylinder, namely: 

$$\tau_c = \frac{\gamma}{\mid d \gamma/dt \mid} = \frac {9}{8\, \pi}
\frac{\rho}{r_e \, \nu}  = 2 \times 10^{21} \frac{P}{\nu}, \eqno(6) $$

\noindent
where $r_e$ is the classical radius of the electron; for $P=2.3$ ms and
$\nu = 10^{18}$ Hz 
we have  $\tau_c \simeq$ 5 s. These electrons could be accelerated either in the outer gaps or above the 
polar caps. In the latter case, we expect that a copious number of 
electron-positron pairs will reach, after suffering
radiation losses,  the light cylinder with energy high enough for X-ray emission.

An origin of the high-energy X-rays in a narrow outer gap can also explain
the narrow profile (intrinsically $<$ 0.04 phase; Kuiper et al. 1998)
 measured in the ROSAT light curve, as well
as the double peak profile, e.g. for the wide fan-beam geometry proposed by 
Romani (1996). However, an 
additional constraint is needed:  the length of the gap over which the hard 
X-rays are produced should be small. In fact, the very strong curvature of the 
magnetic 
field lines would produce a broader pulse profile for a long gap. However, as
Ho (1989) pointed out, strongly curved magnetic field lines,
as in  millisecond pulsars, enhance the 
production of non-thermal emission, allowing 
 outer gaps shorter than  
those of normal radio pulsars. 

According Kuiper et al. (1999), PSR J0218+4232 
is the most likely counterpart of the high-energy EGRET source 2EG J0220+4232 /
3EG J0222+4253 for energies between 100 MeV and 1 GeV. This  would suggest a cut-off
energy at least four orders of magnitude higher than the maximum energy of
10 keV to which the BeppoSAX data showed evidence for the detection of this 
source. Further high sensitivity observations at hard X-ray energies up to 
the high-energy $\gamma$-rays are therefore particularly relevant for the 
understanding of the physics of this and other millisecond pulsars.

\begin{figure}
\centerline{
\hbox{
\psfig{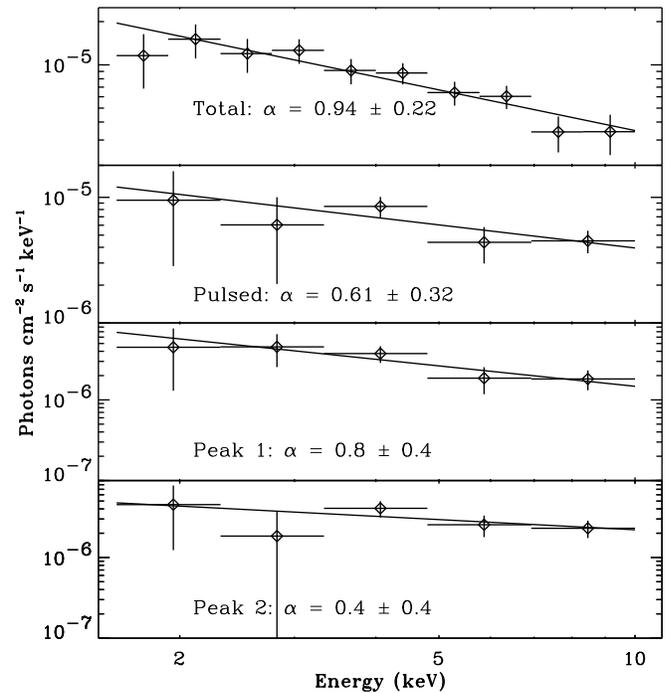}
}}
\caption{\PSR\ spectra as detected by the MECS.
The  lines represent the power--law used as fitting model
}
\end{figure}

\begin{acknowledgements}
\small
TM acknowledges G.Vizzini for his 
technical support on data handling.

\end{acknowledgements}

\end{document}